\documentclass[lettersize,journal]{IEEEtran}
\usepackage{amsmath,amsfonts}
\usepackage{algorithm}
\usepackage{algpseudocode}
\usepackage{array}
\usepackage[caption=false,font=normalsize,labelfont=sf,textfont=sf]{subfig}
\usepackage{textcomp}
\usepackage{stfloats}
\usepackage{url}
\usepackage{verbatim}
\usepackage{graphicx}
\usepackage{cite}
\usepackage{tabularx}
\usepackage{multirow}
\usepackage{adjustbox}
\usepackage{authblk}

\newcolumntype{Y}{>{\centering\arraybackslash}X}

\begin{document}

\title{Fed-NDIF: A Noise-Embedded Federated Diffusion Model For Low-Count Whole-Body PET Denoising}

\author[a]{Yinchi Zhou}
\author[a]{Huidong Xie}
\author[a]{Menghua Xia}
\author[a]{Qiong Liu}
\author[c]{Bo Zhou}
\author[a]{Tianqi Chen}
\author[a]{Jun Hou}
\author[a]{Liang Guo}
\author[a]{Xinyuan Zheng}
\author[d]{Hanzhong Wang}
\author[d]{Biao Li}
\author[e]{Axel Rominger}
\author[e,f]{Kuangyu Shi}
\author[a,b]{Nicha C. Dvornek}
\author[a,b]{Chi Liu}
\affil[a]{Department of Biomedical Engineering, Yale University, New Haven, CT, USA}
\affil[b]{Department of Radiology and Biomedical Imaging, Yale School of Medicine, New Haven, CT, USA}
\affil[c]{Department of Radiology, Northwestern University, Chicago, IL, USA}
\affil[d]{Department of Nuclear Medicine, Ruijin Hospital, Shanghai Jiao Tong University School of Medicine, Shanghai, China.}
\affil[e]{Department of Nuclear Medicine, Inselspital, Bern University Hospital, University of Bern, Bern, Switzerland}
\affil[f]{Computer Aided Medical Procedures and Augmented Reality, Institute of Informatics I16, Technical University of Munich, Munich, Germany}


\markboth{Journal of \LaTeX\ Class Files,~Vol.~14, No.~8, August~2021}%
{Shell \MakeLowercase{\textit{et al.}}: A Sample Article Using IEEEtran.cls for IEEE Journals}


\maketitle

\begin{abstract}
Low-count positron emission tomography (LCPET) imaging can reduce patients' exposure to radiation but often suffers from increased image noise and reduced lesion detectability, necessitating effective denoising techniques. Diffusion models have shown promise in LCPET denoising for recovering degraded image quality. However, training such models requires large and diverse datasets, which are challenging to obtain in the medical domain. To address data scarcity and privacy concerns, we combine diffusion models with federated learning -- a decentralized training approach where models are trained individually at different sites, and their parameters are aggregated on a central server over multiple iterations. The variation in scanner types and image noise levels within and across institutions poses additional challenges for federated learning in LCPET denoising. In this study, we propose a novel noise-embedded federated learning diffusion model (Fed-NDIF) to address these challenges, leveraging a multicenter dataset and varying count levels. Our approach incorporates liver normalized standard deviation (NSTD) noise embedding into a 2.5D diffusion model and utilizes the Federated Averaging (FedAvg) algorithm to aggregate locally trained models into a global model, which is subsequently fine-tuned on local datasets to optimize performance and obtain personalized models. Extensive validation on datasets from the University of Bern, Ruijin Hospital in Shanghai, and Yale-New Haven Hospital demonstrates the superior performance of our method in enhancing image quality and improving lesion quantification. The Fed-NDIF model shows significant improvements in PSNR, SSIM, and NMSE of the entire 3D volume, as well as enhanced lesion detectability and quantification, compared to local diffusion models and federated UNet-based models.
\end{abstract}


\section{Introduction}

\begin{figure*}
    \centering
    \includegraphics[width=\linewidth]{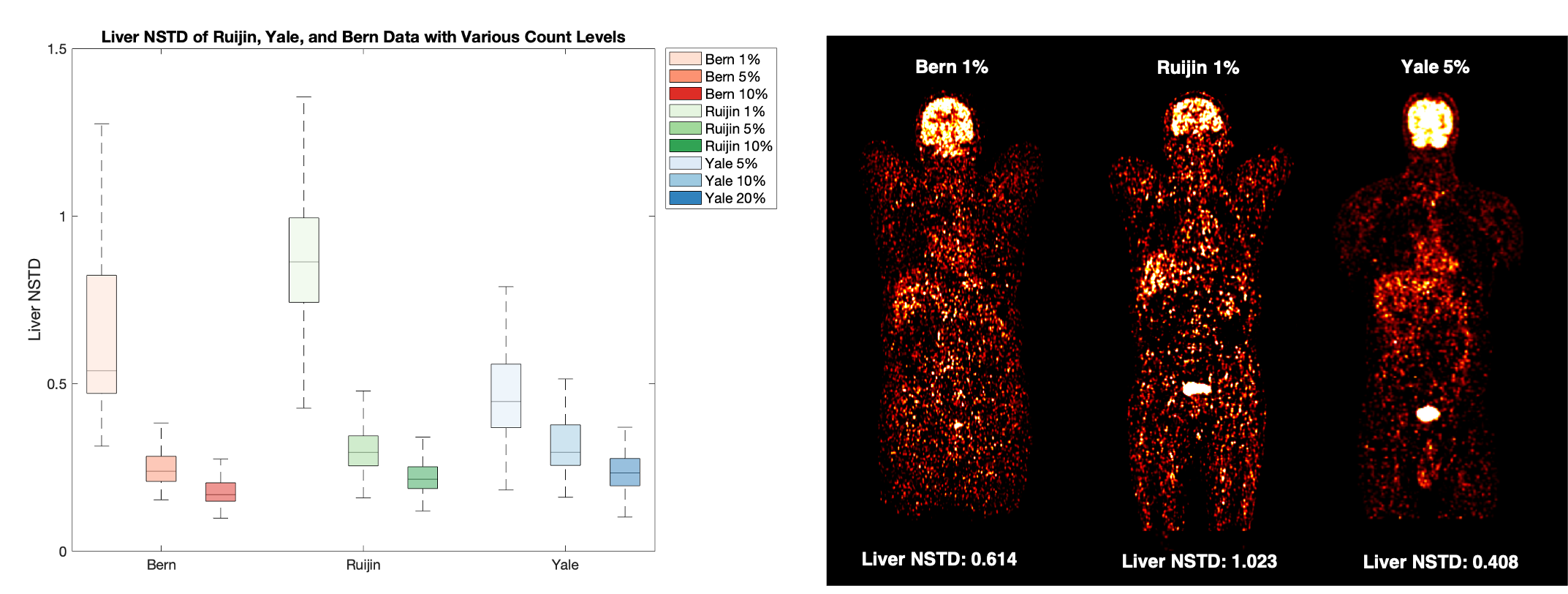}
    \caption{The image noise level is determined by NSTD of a 2 cm x 2 cm x 2 cm cube within the liver region. The liver NSTD distribution for each count level across different datasets is shown on the left. On the right, three example images from each dataset are displayed, illustrating the variation in image noise across different count levels and datasets.}
    \label{nstd}
\end{figure*}

Positron Emission Tomography (PET) has been extensively used in clinical practice to visualize the metabolic and biochemical functions of human tissues and organs, facilitating the detection and diagnosis of abnormalities, including cancer, heart disease, and brain disorders \cite{rohren2004clinical}. A PET scan injects a standardized dose of radioactive tracers (e.g., \textsuperscript{18}F-FDG, \textsuperscript{82}Rb) into the bloodstream and calculates the quantitative uptake value of the radioactive tracer in different organs to reveal lesions. However, the high radiation exposure from the radiotracer increases the risk of genetic mutations and cancer, making dose reduction a desirable step in PET imaging \cite{chawla2010estimated,nievelstein2012radiation,huang2009whole}. Nonetheless, the major drawback of low-count PET (LCPET) includes higher image noise and reduced lesion quantification accuracy and detectability. Therefore, it is crucial to develop methods that can restore the image quality of LCPET to be comparable to full-count PET (FCPET) while reducing the dose level.

Many researchers have successfully applied deep learning algorithms in medical imaging to aid in image processing \cite{isensee2021nnu, ronneberger2015u, yu2023unest}. Convolutional neural networks (CNNs) have shown promise in denoising LCPET \cite{wang20183d, xiang2017deep, liu2022personalized, zhou2021mdpet, xie2023unified, xia2024lpqcm, zhou2020supervised, onishi2021anatomical,ouyang2019ultra, chen20242}. However, CNN-based models have difficulty generalizing to new data especially when there is a gap between the training data and the testing data. This presents challenges for LCPET denoising in real clinical applications, where image noise levels are subject to variability due to differences in scanner types, acquisition protocols, or patients' anatomy \cite{gong2018pet, zhao2020study, liu2021impact, liu2022dose}. Recently, denoising diffusion probabilistic models (DDPMs) and score function-based models \cite{ho2020denoising, sohl2015deep, song2020denoising}, denoted as diffusion models, have demonstrated superior performance in image generation and image translation tasks \cite{saharia2022palette, bansal2024cold, li2023bbdm}. Diffusion models can accurately learn the prior distribution from training data through iterative refinement and therefore generate images with high quality and fidelity. In LCPET image denoising, acquired LCPET images could be used as the condition to guide the generation of FCPET in the reverse process of the diffusion model. For example, \cite{gong2018pet} provided a DDPM with MRI prior information and LCPET to recover information from noisy images for 2D brain PET. \cite{xie2023ddpet} further extended 2D diffusion models to 2.5D to improve the slice consistency of 3D PET images and enhance the quantification of diffusion outputs with a denoised prior from a UNet-based model. \cite{xia2024anatomically} incorporated organ segmentation and lesion prior into the diffusion model training process to improve tumor detectability and preserve anatomical information in PET images.

However, all the aforementioned methods are trained on local datasets and require large-scale labeled training data to achieve satisfying results, which is a fundamental challenge for medical images due to the efforts and costs associated with data acquisition and labeling. Unlike natural images, which can be gathered from open sources to create large datasets such as ImageNet, medical data from different institutions and centers cannot be easily shared to protect patients' privacy \cite{li2020multi, adnan2022federated, sheller2019multi, dayan2021federated}. To address the issues of limited data and privacy concerns, federated learning approaches are proposed to collaborate on data from multiple institutions by exchanging only the model parameters of local site models without exposing individual subjects' information. In a federated learning setting, the individual datasets undergo local training independently in each communication round. The local model parameters are sent to the central server to perform global aggregation, and the global model is subsequently broadcast to all local clients for a new round of local model updates. Through the federated learning scheme, the models are able to learn information from more diverse data without sharing original data, enhancing model performance. For example, on simulated multi-institutional data, \cite{zhou2023fedftn} proposed a federated transferring framework where a global model is pretrained with data from three sites collaboratively in the first stage and then fine-tuned in the second stage. Another federated learning work designed personalized LCPET denoising models for each local site by aggregating only the UNet parameters and keeping feature transformation modules locally trained \cite{zhou2023fedftn}. The resulting federated models in both works outperform the local models for all datasets.

While previous studies have shown the feasibility of deploying federated learning for LCPET denoising with CNN-based models, federated diffusion models remain under-explored. \cite{de2024training} adapted the federated averaging algorithm \cite{mcmahan2017communication} on DDPMs to generate high-quality 2D natural images. For LCPET denoising, \cite{zhou2024feddd} investigated the federated diffusion model by fine-tuning a pre-trained model in a federated manner. However, the study only explored one noise level for each local dataset, and the performance of fine-tuned models heavily relied on the quality of the pre-trained model and the similarity between the pre-trained datasets and the local datasets.

One of the primary challenges in federated learning for LCPET is the heterogeneity in image noise levels both within and across different datasets. Within each dataset, image noise levels are influenced by the number of counts, which are determined by factors such as the injected dose and the duration of image acquisition. During training, we simulate various noise levels by downsampling the FCPET images to specific noise levels (e.g., 1\%, 5\%, 10\%) to reflect different count scenarios encountered in clinical practice. It is observed that image noise levels vary significantly according to the count level, as illustrated in Fig.\ \ref{nstd}. Additionally, across different datasets, even with the same simulated count level, substantial variability in image noise exists due to discrepancies in scanning protocols and scanner sensitivities across different institutions. Consequently, there is a critical need for a denoising approach that can generalize to varying count levels and accurately quantify image noise variations across different institutions. Previous studies have addressed differences in noise levels by either training separate neural networks for each count level or embedding count information into the network for local datasets \cite{xie2023ddpet, li2022noise}. However, these approaches fail to address the issue of image domain noise variability across institutions with different scanning protocols and equipment in a federated learning scheme.

 In our work, we propose a noise-embedded federated diffusion model for LCPET denoising using multicenter datasets and various count levels. Briefly, based on the 2.5D diffusion model for PET denoising framework proposed by \cite{xie2023ddpet}, we incorporate the federated averaging (FedAvg) algorithm into the diffusion training process to train a global model and then fine-tune this global model to optimize denoising performance on local datasets. Additionally, inspired by \cite{liu2022personalized}, we determine the image noise in terms of the normalized standard deviation (NSTD) from the liver region of interest and embed this PET image-based noise information for different count levels and datasets into the diffusion model to address the variation within and across training data. We validate the effectiveness of our method on three whole-body PET datasets with multiple count levels from different imaging sites: the University of Bern (Siemens Quadra scanner), Ruijin Hospital in Shanghai (United Imaging uExplorer scanner), and Yale New Haven Hospital (Siemens mCT scanner). Our main contributions are threefold: 1) We demonstrate the promise of federated learning to improve the performance of diffusion models on LCPET denoising for whole-body PET images. 2) We show that embedding image noise information into the model can further enhance model learning. 3) We evaluate our proposed method on three datasets and demonstrate that the federated diffusion model consistently outperforms local diffusion models and other federated models in terms of image quality and lesion quantification.

\begin{figure*}[htb!]
    \centering
    \includegraphics[width=0.9\textwidth]{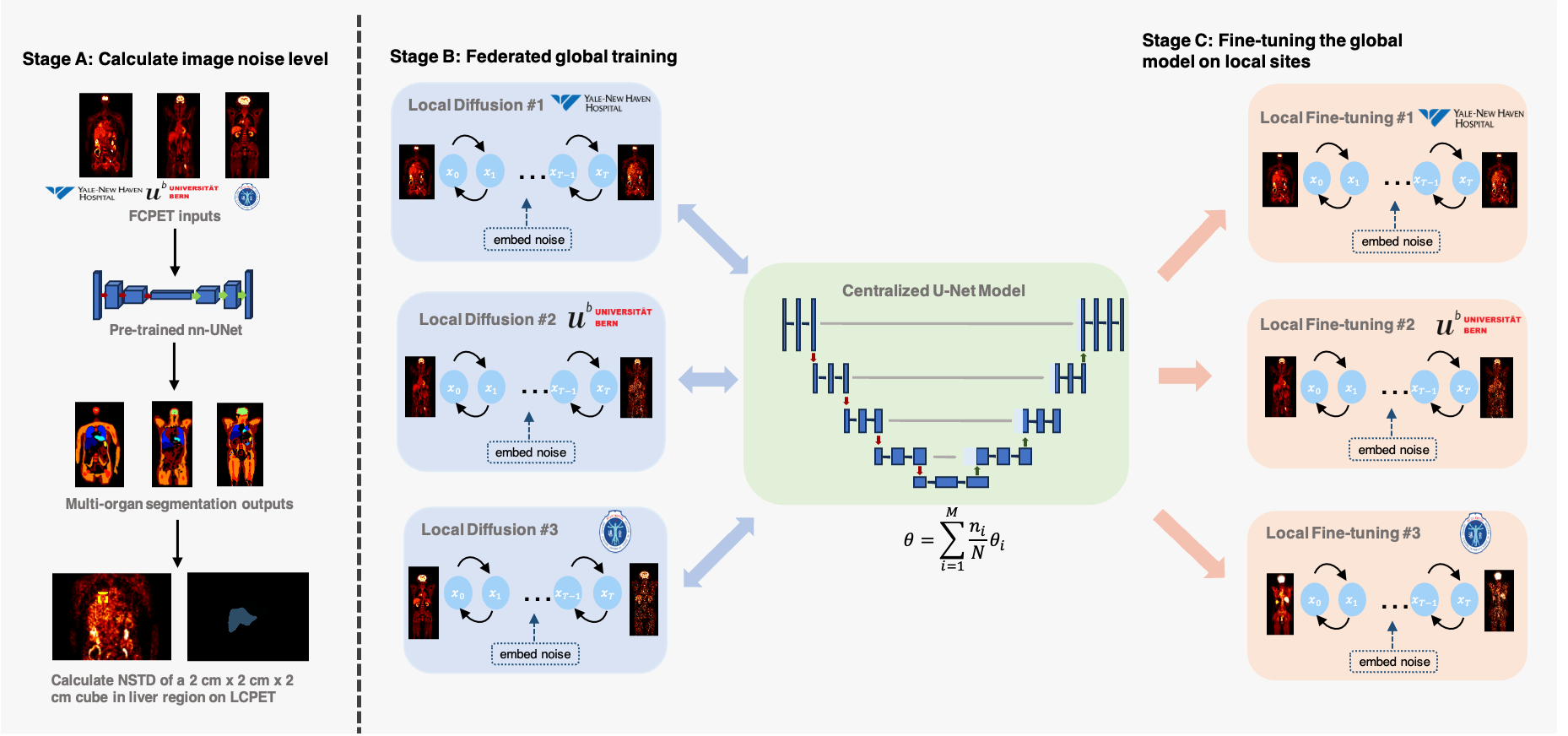}
    \caption{Overview of the proposed Fed-NDIF. In Stage A, the liver NSTD on LCPET is calculated and later embedded into the diffusion model in Stage B. During Stage B, local diffusion models are trained independently at each site. Aggregation occurs every few epochs, wherein model parameters are averaged to update the centralized model. This centralized model is then sent back to the local sites to initiate a new round of local training. In Stage C, the global model is further fine-tuned using local datasets.}
    \label{framework}
\end{figure*}

\section{Materials and Methods}
\subsection{Datasets}
\textbf{Bern Dataset}: This dataset was acquired at the Department of Nulcear Medicine, University of Bern, Switzerland \cite{xue2022cross}, using a Simens Biograph Vision Qiadra total-body PET/CT system. It comprises 134 \textsuperscript{18}F-FDG subjects, with 99 randomly selected for training, 5 for validation, and 30 for testing. The LCPET training data were generated by uniformly down-sampling the PET list-mode data into 1\%, 2\%, 5\%, 10\%, 25\%, and 50\% count levels. The validation and testing datasets were available in 1\%, 5\%, and 10\% count levels. Both LCPET and FCPET were reconstructed using the OSEM (Ordered Subsets Expectation Maximization) algorithm with 4 iterations and 20 subsets. A 5 mm FWHM Gaussian filter was applied post-reconstruction. The reconstructed images have dimensions of 440 x 440 x 644 and a voxel size of 1.65 x 1.65 x 1.65 mm\textsuperscript{3}.      

\textbf{Ruijin Dataset}: This dataset was collected at Ruijin Hospital in Shanghai, China, using a United Imaging uExplorer total-body PET/CT system. A total of 234 subjects injected with \textsuperscript{18}F-FDG tracer were included in this dataset, which was split into 199 subjects for training, 5 for validation, and 30 for testing. The LCPET training data were created by uniformly down-sampling the PET list-mode data into 1\%, 2\%, 5\%, 10\%, 25\%, and 50\% count levels. The validation and testing datasets were downsampled into 1\%, 5\%, and 10\% count levels. Both LCPET and FCPET were reconstructed using the OSEM algorithm with 4 iterations and 20 subsets. A 5 mm FWHM Gaussian filter was applied after the reconstructions. The reconstructed images have an image size of 360 x 360 x 674 and a voxel size of 1.67 x 1.67 x 2.89 mm\textsuperscript{3}.

\textbf{Yale Dataset}: This dataset was obtained at Yale-New Haven Hospital, USA, using a Siemens Biograph mCT scanner. It comprises 135 subjects scanned with \textsuperscript{18}F-FDG, with 100 subjects used for training, 5 for validation, and 30 for testing. The LCPET (Low Count PET) training data were produced by downsampling the PET list-mode data to 5\%, 10\%, 20\%, 30\%, and 50\% count levels. For the validation and testing subjects, images at 5\%, 10\%, and 20\% count levels are available. All PET images were reconstructed using the OSEM algorithm with 2 iterations and 21 subsets. Additionally, a 5 mm FWHM Gaussian filter was applied to the reconstructed images. The images have dimensions of 400 x 400 in the transverse plane, while the axial dimension varies depending on the patients' height. The voxel size is 2.04 x  2.04 x 2.03 mm\textsuperscript{3}. Among the 30 testing subjects, 15 have lesion segmentation labels provided by clinicians.

\subsection{Framework Overview}
The proposed Federated Diffusion Training Framework (Fed-NDIF) consists of three stages: image noise level calculation, federated global training, and local fine-tuning, as illustrated in Fig.\ \ref{framework}. In Stage A, the image noise level for LCPET is determined by calculating the normalized standard deviation (NSTD) from a 2 cm x 2 cm x 2 cm cube within a relatively uniform liver region, which is segmented using a pre-trained nn-UNet model on FCPET \cite{salimi2022deep}. In Stage B, a global 2.5D guided diffusion model is trained with three datasets using a federated learning approach, conditioned by both low-count and full-count images, with the liver NSTD information embedded into the model. The federated global training process spans 30 iterations, each consisting of 10,000 steps. During each iteration, individual diffusion models are trained and refined on local datasets, and model parameters are aggregated using the FedAvg algorithm \cite{mcmahan2017communication} at the end of each iteration. The aggregated parameters are then redistributed to the local sites for subsequent rounds of training. In Stage C, the averaged global models are fine-tuned locally on the three datasets.

\subsection{Conditional 2.5D Diffusion Models for PET Denoising}
Diffusion models are a class of generative models that learn the target data distribution and then allow us to synthesize new samples from the learned distribution. The diffusion model has a forward process and a reverse process built on the Markov chain process. In the forward process, Gaussian noise is gradually added to the target image $x_0 \sim q(x_0)$ over $T$ time steps, producing latent variables $x_1, x_2, ..., x_T$ defined as follow:
\begin{align}
    &q(x_1,...,x_T|x_0) = \prod_{t=1}^T q(x_t|x_{t-1}), \\
    q(x_t |x_{t-1}) = &\mathcal{N}(x_t;\sqrt{1-\beta_t}x_{t-1}, \beta_t\mathbf{I}), \forall t \in {1,...,T}
\end{align}
where $T$ and $\beta_1, ..., \beta_T \in [0,1)$ represent the number of diffusion steps and the variance schedule across these steps. $\mathbf{I}$ is the identity matrix and $\mathcal{N}(x;\mu,\sigma)$ represents the Gaussian distribution with mean $\mu$ and variance $\sigma$. Denote $\alpha_t = 1-\beta$, $\bar{\alpha} = \prod_{s=0}^t\alpha_s$, and $\epsilon\sim\mathcal{N}(0,I)$, the Markov property allows us to sample any arbitrary step conditioned on $x_0$. The following equations are obtained:
\begin{align}
    q(x_t|x_0) = \mathcal{N}(x_t;\sqrt{\bar{\alpha_t}}x_0, (1-\bar{\alpha_t})\mathbf{I}),& \\
    x_t = \sqrt{\bar{\alpha_t}}x_0 + (1-\bar{\alpha_t})\epsilon &
\end{align}
With a sufficiently large $T$ and appropriate variance scheduling $\beta_t$, $x_T$ becomes a nearly isotropic Gaussian. In the reverse process, the posterior $q(x_{t-1}|x_t)$ is also Gaussian. We can then obtain a sample from $q(x_0)$. However, because $q(x_0)$ is unknown, we need to approximate the distribution $q(x_{t-1}|x_t)$ by a network with parameters $\theta$, and the parameterized distribution is as follow:
\begin{equation}
    p_\theta(x_{t-1}|x_t) = \mathcal{N}(x_{t-1};\mu_\theta(x_t,t),{\sigma_t}^2\mathbf{I})
\end{equation}
\cite{ho2020denoising} found training a model to predict $\epsilon$ with the parametrized $\epsilon_\theta(x_t,t)$ gives better results than parameterizing $\mu_\theta(x_t,t)$. Note, we have adapted the conditional diffusion models for PET denoising proposed by \cite{gong2018pet} and \cite{xie2023ddpet}, where the network is conditioned by LCPET images $x_{LC}$ as the additional input, so the estimated posterior becomes 
\begin{equation}
\label{eq:mu}
    \mu_\theta(x_t,t,x_{LC}) = \frac{1}{\sqrt{\alpha_t}}(x_t-\frac{1-\alpha_t}{\sqrt{1-\bar{\alpha}_t}}\epsilon_\theta(x_t,t,x_{LC}))
\end{equation}
The training objective of $\epsilon_\theta(x_t,t)$ can be formulated as
\begin{equation}
\label{eq:train}
    \mathbb{E}_{t,x,\epsilon}[{||\epsilon-\epsilon_\theta(x_t,t,x_{LC})||}^2],
\end{equation}
In inference, the iterative step to obtain $x_0$ from $x_T$ is written as 
\begin{equation}
    x_{t-1} = \mu_\theta(x_t,t,x_{LC}) + {\sigma_t}^2\mathbf{I}
\end{equation}
The diffusion model in our experiments is based on the 2.5D conditional diffusion model proposed by \cite{xie2023ddpet}. Specifically, $x_{LC}$ consists of 31 neighboring low-count slices, and $x_0$ is the corresponding central full-count slice. We adopted the following modifications proposed for PET denoising diffusion models during inference by Xie et al.: 1) the starting Gaussian noise in the reverse process is fixed when reconstructing all slices of the entire 3D volume to address slice inconsistency; 2) two different Gaussian noise variables $\epsilon_0$ and $\epsilon_1$ are initialized for $x_T$ as the starting noise, and $x_{T-1}$  is acquired by averaging the results of $x_{T-1}$ for two noise variables 3) a UNet denoised prior with the initialized Gaussian noise $\epsilon_0$ and $\epsilon_1$ is used as $x_T$. The details for the inference algorithm are discussed in their paper \cite{xie2023ddpet}.
\subsection{Noise Embedding}
To make the diffusion model generalize better to different count levels and institutions, the image noise level for LCPET is embedded into the diffusion model. To determine the image noise level, a pre-trained nnUNet model is used to segment FCPET to obtain organ labels. The liver label is then used to localize the liver region on LCPET, and the liver NSTD is calculated in a 2 cm x 2 cm x 2 cm cube in the relatively uniform area of the liver. The liver NSTD is embedded with a linear layer and added as an additional input to the neural network during the reverse process, where the parameterized $\epsilon_\theta(x_t,t,x_{\mathrm{LC}})$ becomes $\epsilon_\theta(x_t,t,x_{\mathrm{LC}}, noise_{\mathrm{LC}})$ in Eq.\ \ref{eq:mu} and \ref{eq:train}.

\subsection{Federated Learning}

In the federated global training stage, the centralized model aggregates model parameters from local models using the Federated Averaging algorithm (FedAvg) \cite{mcmahan2017communication}. The federated averaging involves local training with $K$ steps and global averaging over $C$ rounds for $M$ clients. In local training, the local diffusion model takes gradient descent steps to optimize Eq.\ \ref{eq:train} with the embedded image noise level. After training local models with $K$ steps, the neural network parameters $\theta$ are aggregated across all datasets based on the sample sizes $n_i$ of each local dataset, where $\theta_{glb} = \sum_{i=1}^{M}\frac{n_i}{N}\theta_i$, $N$ is the total number of samples across all sites. The global model is then sent back to the local datasets for a new round of training. The pseudocode for the federated diffusion training is shown in Algorithm \ref{alg_fed}.

\begin{algorithm}[t]
\caption{Federated diffusion training algorithm}\label{alg_fed}
\begin{algorithmic} 
\State \textbf{Require:} dataset $D=\{D_1, D_2, ..., D_M\}$ for $M$ clients, the model parameters $\theta$, the number of local training steps $K$, the number of communication round $C$, and diffusion steps $T$
\State \textbf{Initialize $\theta_0$}
\For{$c=0,1,..,C$}
\State send $\theta_c$ to all clients for local diffusion training
    \For{each client $i \in M$ in parallel, repeat $K$ steps}
        \State $x_0 \sim q(x_0)$ \Comment{Sample target full-count slices}
        \State $x_{LC}$, $noise_{x_{\mathrm{LC}}}$ \\ \hfill \Comment{paired low-count slices and 
        image noise}
        \State $t \sim \mathrm{Uniform}({1,..,T})$ \Comment{sample diffusion time steps}
        \State $\epsilon \sim \mathcal{N}(0,\textbf{I})$ \Comment{Sample added noise}
        \State $\nabla_\theta||{\epsilon}-\epsilon_\theta(\sqrt{\bar{\alpha}_t}x_0+\sqrt{1-\bar{\alpha}_t}{\epsilon},t,x_{\mathrm{LC}}, noise_{\mathrm{LC}})$ \\ \hfill \Comment{Take gradient descent step}
    \EndFor
    \State $\theta_{c+1} = \sum_{i=1}^M\frac{n_i}{N}\theta_{c}^{i}$ \Comment{Average local model parameters}
\EndFor

\end{algorithmic}
\end{algorithm}

\subsection{Evaluation Metrics and Baseline Comparison}
To evaluate the denoising performance of the proposed method, we assessed the LCPET denoising outputs in terms of image quality metrics and lesion quantification. For image quality, we used the Peak Signal-to-Noise Ratio (PSNR), Structural Similarity Index (SSIM), and Normalized Mean Square Error (NMSE) between the entire 3D volume of the denoised outputs and full-count images \cite{kollem2019review}. To evaluate the denoising performance in critical pathological regions, we calculated the NMSE and the relative percentage error of the maximum and mean Standard Uptake Value (SUV\textsubscript{max} and SUV\textsubscript{mean}) on individual lesions for the subjects with ground truth lesion segmentation labels using the following formula:
\begin{equation}
    \mathrm{Percentage\ Error} = \frac{\sum_{i\in L}|Y_i^{pred}-Y_i^{gt}|}{\sum_{i\in L}Y_i^{gt}} * 100\%
\end{equation}
where $L$ represents lesion regions, $Y^{pred}$ is the intensity value in the lesion regions of the denoised outputs, and $Y^{gt}$  is the intensity value in the lesion regions of the FCPET. We compared our model against local diffusion models and a federated UNet model. The local diffusion models include the vanilla denoising diffusion implicit models with 25 time steps (DDIM-PET) for PET denoising \cite{gong2024pet} and the diffusion model with fixed Gaussian noise and denoised prior in the reverse process (DDPET) \cite{xie2023ddpet}. Note that our proposed method is based on DDPET, incorporating federated learning and noise embedding. For UNet-based models, we compared against the UNet with FedAvg (Fed-UNet) and the personalized federated UNet model for PET denoising (FedFTN) \cite{zhou2023fedftn}.

\begin{figure*}[htb!]
    \centering
    \includegraphics[width=0.9\linewidth]{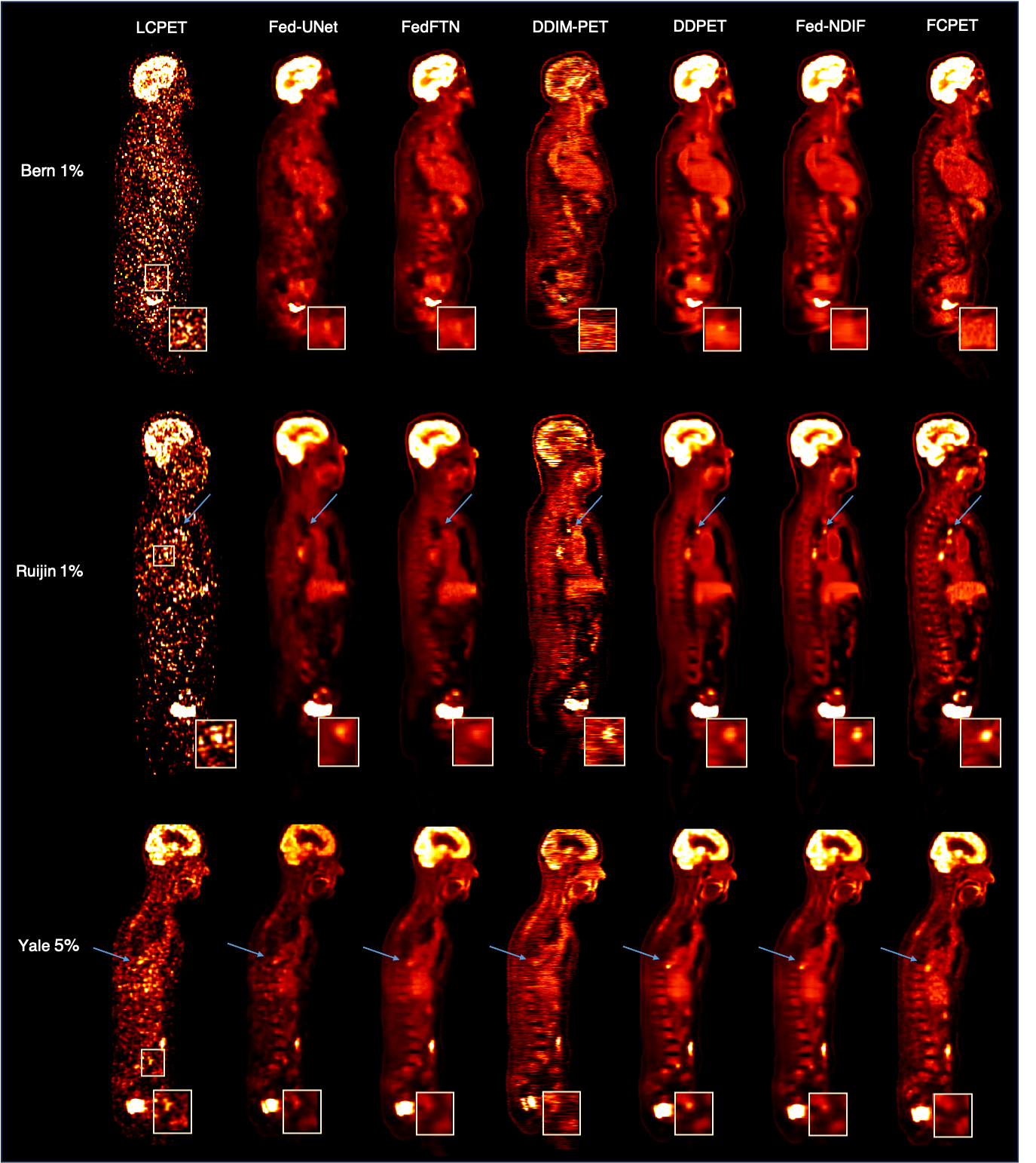}
    \caption{Sample images from three institutions with the lowest count levels are shown. The yellow boxes are cropped regions that are magnified and displayed at the bottom-right corner of the image. The blue arrows indicate the lesions on FCPET. Federated diffusion models produce images with higher resolution and fewer false-positive and false-negative lesions, as indicated by blue arrows and yellow boxes.}
    \label{comparison}
\end{figure*}

\subsection{Implementation Details}
We implemented our method in PyTorch and performed experiments using an NVIDIA A40 GPU with 48 GB of memory. In the preprocessing step, the outermost background areas of PET images are removed. In the federated global training stage, the number of local training steps was set to 10,000. We used a batch size of 6 and trained the federated diffusion model for 30 communication rounds. The model was optimized using the Adam optimizer with a learning rate of 1e-4 and a cosine scheduler. The federated global model was fine-tuned on the local dataset with a reduced learning rate of 1e-5 for 10 epochs.

\begin{table*}[htb!]
    \centering
    \scriptsize
    \caption{Evaluation metrics for different denoising methods on three count levels for three datasets (mean ± std), including 30 subjects. The best results are bolded.}
    \label{table1}
    \begin{adjustbox}{width=\textwidth}
        \begin{tabularx}{\textwidth}{c|ccY|ccY|ccY}
            \hline
            \multicolumn{10}{c}{Bern} \\ 
            \hline
            \multirow{2}{*}{Method} & \multicolumn{3}{c|}{1\%} & \multicolumn{3}{c|}{5\%} & \multicolumn{3}{c}{10\%} \\ \cline{2-10}
             & PSNR $\uparrow$ & SSIM $\uparrow$ & NMSE $\downarrow$ & PSNR $\uparrow$ & SSIM $\uparrow$& NMSE $\downarrow$ & PSNR$\uparrow$ & SSIM $\uparrow$ & NMSE $\downarrow$ \\ \hline 
            LC PET & 47.546 ± 6.235 & .863 ± .098 & .483 ± .171 & 55.031 ± 5.840 & .960 ± .024 & .196 ± .040 & 57.696 ± 5.796 & .972 ± .012 & .144 ± .030 \\ 
            Fed-UNet & 52.394 ± 5.390 & .967 ± .019 & .269 ± .048 & 54.719 ± 5.401 & .979 ± .006 & .204 ± .014 & 55.270 ± 5.411 & .978 ± .004 & .191 ± .011 \\ 
            FedFTN &  54.196 ± 5.367 & .980 ± .012  & .214 ± .038 & 56.228 ± 5.444 & .989 ± .006 & .168 ± .016 & 56.792 ± 5.492 & .991 ± .005 & .157 ± .014  \\
            DDIM-PET & 43.910 ± 5.139 & .925 ± .041 & .693 ± .035 & 43.925 ± 5.406 & .904 ± .051 & .691 ± .052 & 43.963 ± 5.406 & .878 ± .062 & .688 ± .055 \\
            DDPET & 55.242 ± 5.139 & .979 ± .013 & .190 ± .035 & 57.162 ± 5.258 & .985 ± .007 & .152 ± .023 & 59.305 ± 5.376 & .988 ± .004 & .119 ± .024 \\ 
            \textbf{Fed-NDIF} & \textbf{55.799 ± 5.203}\textsuperscript{\textdagger} & \textbf{.983 ± .011}\textsuperscript{\textdagger} & \textbf{.179 ± .037}\textsuperscript{\textdagger} & \textbf{59.266 ± 5.371}\textsuperscript{\textdagger} & \textbf{.989 ± .005}\textsuperscript{\textdagger} & \textbf{.120 ± .022}\textsuperscript{\textdagger} & \textbf{60.615 ± 5.568}\textsuperscript{\textdagger} & \textbf{.989 ± .004}\textsuperscript{\textdagger} & \textbf{.103 ± .023}\textsuperscript{\textdagger} \\ \hline \hline
            \multicolumn{10}{c}{Ruijin} \\ 
            \hline
            \multirow{2}{*}{Method} & \multicolumn{3}{c|}{1\%} & \multicolumn{3}{c|}{5\%} & \multicolumn{3}{c}{10\%} \\ \cline{2-10}
             & PSNR $\uparrow$ & SSIM $\uparrow$ & NMSE $\downarrow$ & PSNR $\uparrow$ & SSIM $\uparrow$& NMSE $\downarrow$ & PSNR$\uparrow$ & SSIM $\uparrow$ & NMSE $\downarrow$ \\ \hline
            LC PET & 42.685 ± 7.218 & .771 ± .130 & .586 ± .130 & 51.720 ± 6.869 & .942 ± .035 & .207 ± .046 & 54.632 ± 6.775 & .963 ± .018 & .149 ± .036 \\
            Fed-UNet & 49.474 ± 6.382 & .954 ± .028 & .269 ± .036 & 52.775 ± 6.065 & .976 ± .011 & .184 ± .026 & 53.540 ± 5.942 & .978 ± .008 & .168 ± .022 \\
            FedFTN & 50.703 ± 6.108 & .967 ± .021 & .229 ± .032 & 54.269 ± 5.815 & .984 ± .010 & .152 ± .027 & 55.389 ± 5.653 & .988 ± .001 & .134 ± .024 \\
            DDIM-PET & 40.068 ± 5.829 & .871 ± .063 & .772 ± .038 & 40.265 ± 5.795 & .861 ± .048 & .755 ± .040 & 40.319 ± 5.797 & .836 ± .074 & .750 ± .041 \\
            DDPET & 50.796 ± 6.003 & .968 ± .019 & .226 ± .034 & 55.306 ± 6.060 & .983 ± .009 & .136 ± .031 & 56.896 ± 5.955 & .984 ± .008 & .114 ± .031 \\ 
            \textbf{Fed-NDIF} & \textbf{51.316 ± 6.198}\textsuperscript{\textdagger} & \textbf{.970 ± .019}\textsuperscript{\textdagger} & \textbf{.214 ± .028}\textsuperscript{\textdagger} & \textbf{55.723 ± 6.225}\textsuperscript{\textdagger} & \textbf{.984 ± .009}\textsuperscript{\textdagger} & \textbf{.131 ± .031}\textsuperscript{\textdagger} & \textbf{57.264 ± 6.168}\textsuperscript{\textdagger} & \textbf{.985 ± .009}\textsuperscript{\textdagger} & \textbf{.109 ± .027}\textsuperscript{\textdagger} \\ \hline \hline
            \multicolumn{10}{c}{Yale} \\
            \hline
            \multirow{2}{*}{Method} & \multicolumn{3}{c|}{1\%} & \multicolumn{3}{c|}{5\%} & \multicolumn{3}{c}{10\%} \\ \cline{2-10}
             & PSNR $\uparrow$ & SSIM $\uparrow$ & NMSE $\downarrow$ & PSNR $\uparrow$ & SSIM $\uparrow$& NMSE $\downarrow$ & PSNR$\uparrow$ & SSIM $\uparrow$ & NMSE $\downarrow$ \\ \hline
            LC PET & 48.459 ± 6.527 & .870 ± .083 & .380 ± .110 & 51.881 ± 6.494 & .922 ± .052 & .256 ± .073 & 55.519 ± 6.526 & .956 ± .028 & .168 ± .048 \\ 
            Fed-UNet & 47.566 ± 5.668 & .912 ± .038 & .405 ± .028 & 51.705 ± 5.726 & .957 ± .022 & .252 ± .027 & 52.784 ± 5.655 & .966 ± .014 & .222 ± .016 \\ 
            FedFTN & 54.192 ± 5.714 & .966 ± .023 & .192 ± .036 & 55.695 ± 5.626 & .974 ± .017 & .161 ± .028 & 57.183 ± 5.600  & .981 ± .013 & .135 ± .022 \\
            DDIM-PET & 44.035 ± 5.177 & .883 ± .067 & .611 ± .067 & 44.403 ± 5.116 & .897 ± .056 & .587 ± .075 & 44.702 ± 5.086 & .905 ± .047 & .569 ± .083 \\
            DDPET & 52.600 ± 6.027 & .957 ± .029 & .231 ± .051 & 54.085 ± 5.852 & .968 ± .020 & .194 ± .037 &  55.418 ± 5.802 & .975 ± .015 & .167 ± .033 \\ 
            Fed-NDIF & \textbf{54.337 ± 6.075}\textsuperscript{\textdagger} & \textbf{.968 ± .024}\textsuperscript{\textdagger} & \textbf{.190 ± .043}\textsuperscript{\textdagger} & \textbf{56.364 ± 6.091}\textsuperscript{\textdagger} & \textbf{.977 ± .017}\textsuperscript{\textdagger} & \textbf{.150 ±.032}\textsuperscript{\textdagger} & \textbf{58.547 ± 6.243}\textsuperscript{\textdagger} & \textbf{.984 ± .013}\textsuperscript{\textdagger} & \textbf{.117 ± .026}\textsuperscript{\textdagger} \\ \hline
        \end{tabularx}
    \end{adjustbox}
    \begin{minipage}{\textwidth}
    \vspace{0.1cm}
    \scriptsize \textdagger \ $p < 0.005$ between DDPET and Fed-NDIF based on Wilcoxon rank test. 
    \end{minipage}
\end{table*}

\begin{figure*}[htb!]
    \centering
    \includegraphics[width=0.95\linewidth]{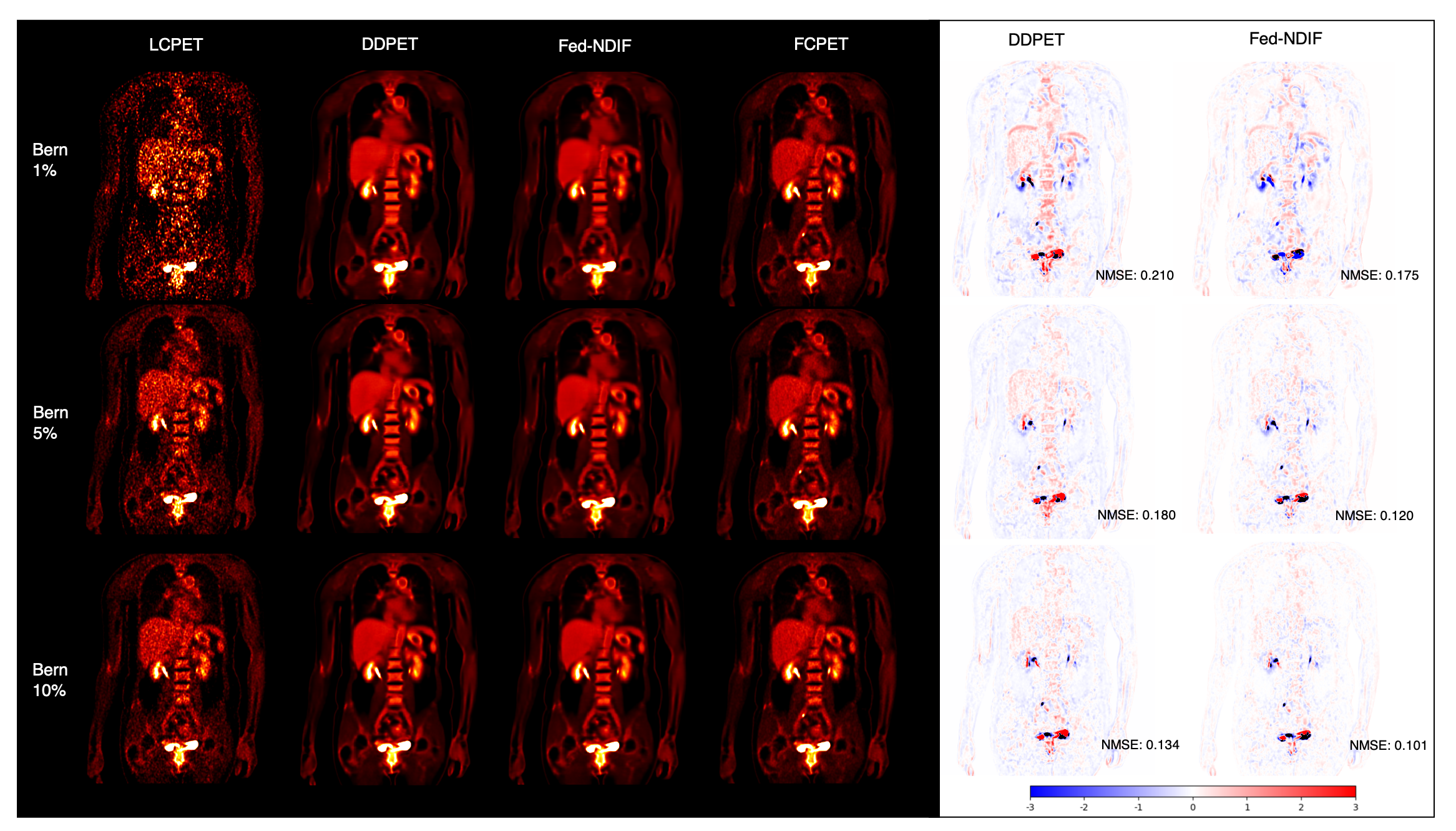}
    \caption{Comparison of the local diffusion model and the federated diffusion model on 1\%, 5\%, 10\% count levels for Bern data. Images are normalized to standard count. The left shows the denoised images, and the right shows the absolute error map of the two methods. Fed-NDIF has reduced the overestimation in liver and cardiac region and underestimation of the body. }
    \label{bern}
\end{figure*}

\begin{figure*}[htb!]
    \centering
    \includegraphics[width=0.9\linewidth]{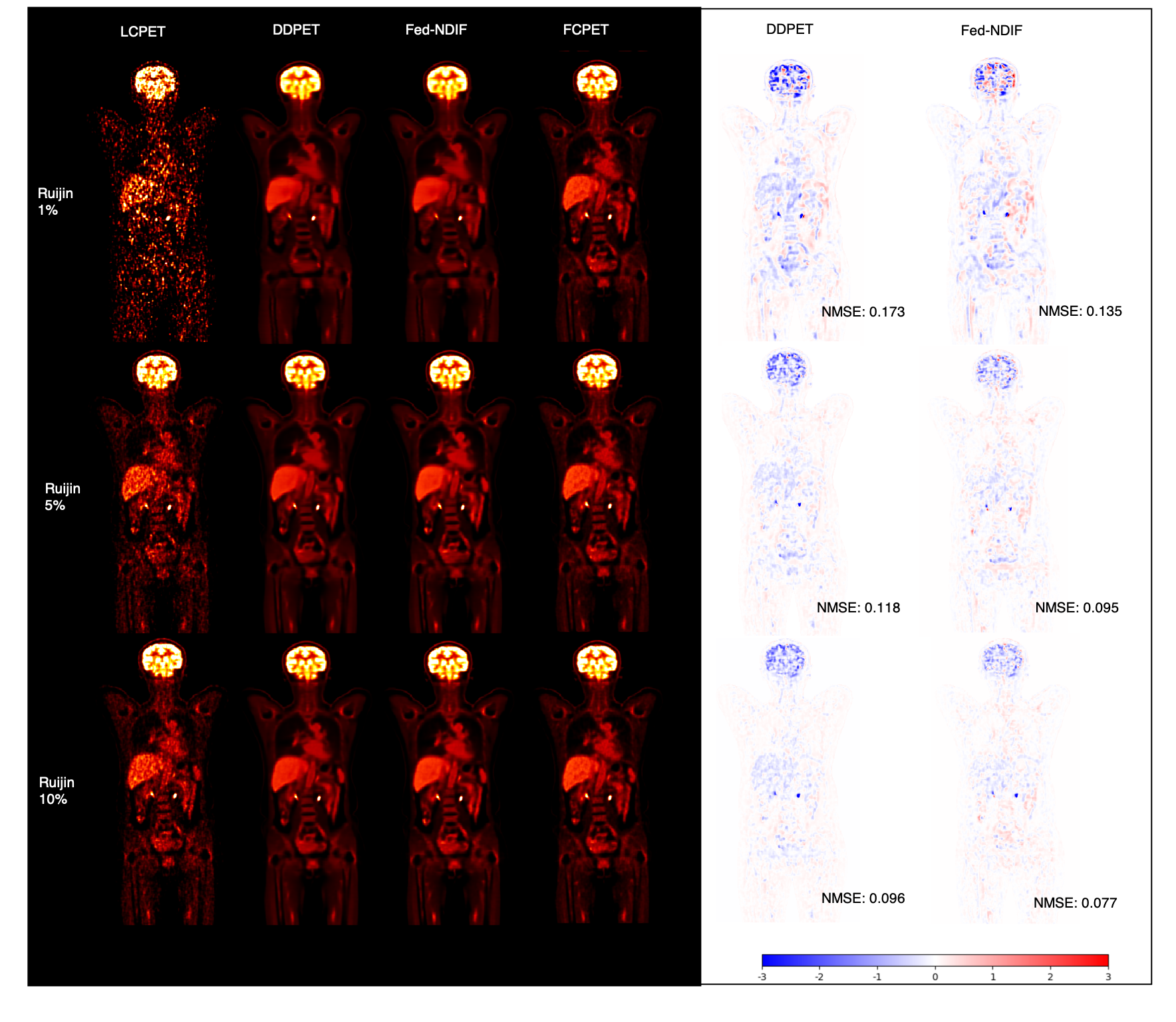}
    \caption{Comparison of the local diffusion model and the federated diffusion model on 1\%, 5\%, 10\% count levels for Ruijin data. Images are normalized to standard count. Fed-NDIF has reduced the overestimation and underestimation of the denoised outputs. }
    \label{ruijin}
\end{figure*}

\begin{figure*}[htb!]
    \centering
    \includegraphics[width=0.95\linewidth]{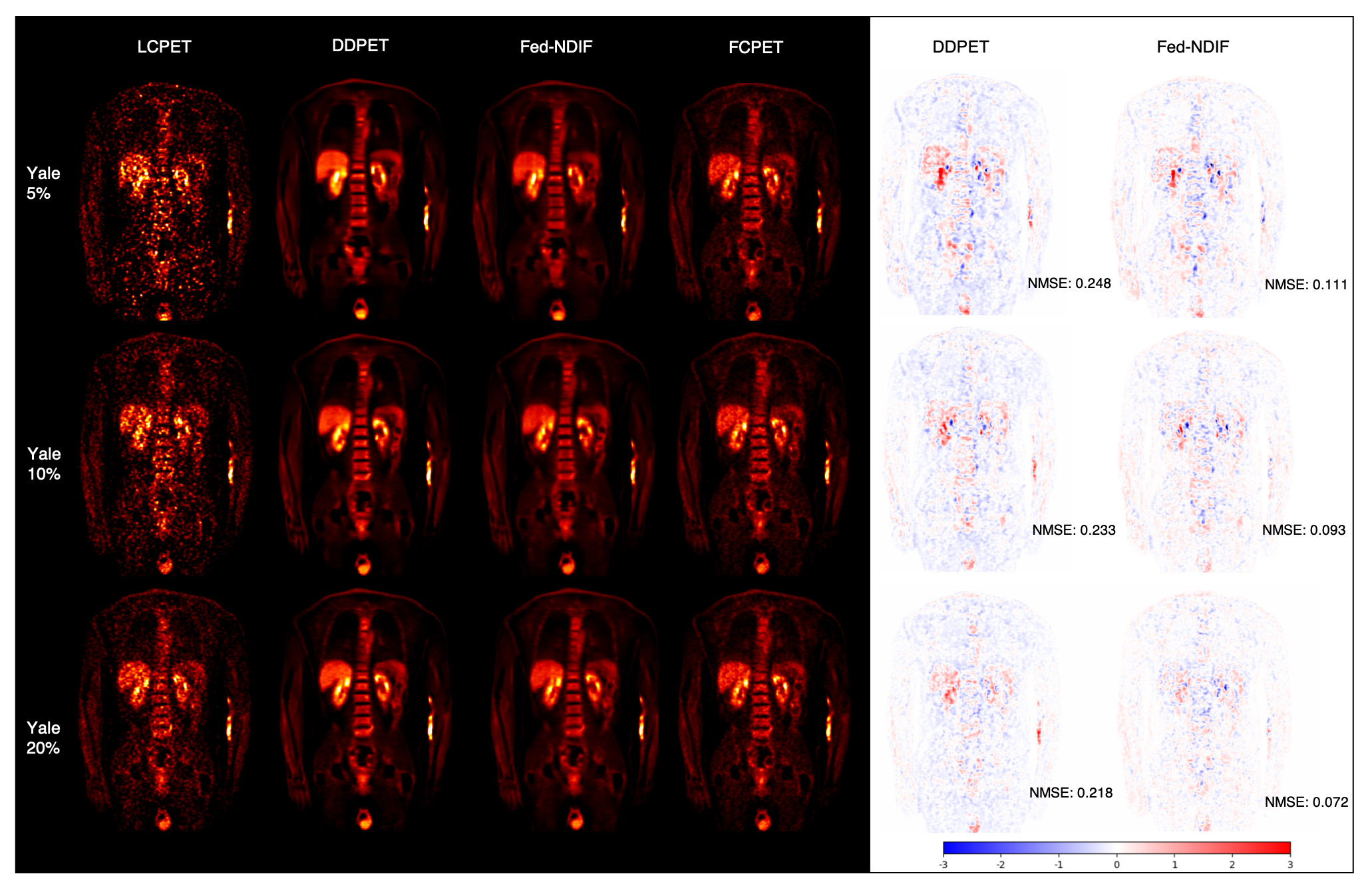}
    \caption{Comparison of the local diffusion model and the federated diffusion model on 5\%, 10\%, 20\% count levels for Yale data. Images are normalized to total counts in FCPET. Similarly, Fed-NDIF has reduced the overestimation in liver and kidney region and the underestimation in body. }
    \label{yale}
\end{figure*}

\section{Results}

\subsection{Baseline comparison}
The image quality of the denoised outputs was evaluated at multiple count levels for three datasets. Table \ref{table1} summarizes the quantitative results of the baseline methods and the proposed Fed-NDIF. Fed-NDIF outperformed both federated UNet-based models and local diffusion models at all count levels and datasets in terms of PSNR, SSIM, and NMSE. Compared with the local diffusion model DDPET, where our method added federated learning and noise embedding, Fed-NDIF resulted in significantly higher NMSE, PSNR, and SSIM ($p < 0.005$). Specifically, for the Bern dataset, NMSE improved by 12.8\% and PSNR improved by 3.8\%, averaged across all count levels. For the Ruijin dataset, NMSE improved by 4.8\%, and PSNR improved by 0.8\%, averaged across all count levels. For the Yale dataset, NMSE improved by 22.8\%, and PSNR improved by 4.4\%, averaged across all count levels.

Fig.\ \ref{comparison} shows the visual comparison of a sample subject from the lowest count level of each dataset using different denoising approaches. The proposed federated diffusion model Fed-NDIF can recover PET images with extremely low count levels, such as 1\% and 5\%, and very noisy images. We found that diffusion-based models have higher overall image resolution than federated UNet-based models. From the sagittal view, it is clearly seen that diffusion models can better recover the structures and fine details of organs, especially in the aorta wall and the spine. For diffusion models, DDIM-PET has worse quantification in brain regions, possibly because it did not use a denoised prior from the UNet model. Artifacts between slices were also observed in DDIM-PET when using different Gaussian noise as the starting point in the reverse process. Many false positives and false negatives were corrected by the federated diffusion model Fed-NDIF. For the Bern dataset, when comparing DDPET and Fed-NDIF, Fed-NDIF reduces the false positive prediction in the lower abdominal region compared with DDPET. In LCPET, this region resembles a lesion but indeed appears as noise. The removal of this false positive demonstrates the capability of Fed-NDIF to accurately recover the image from very noisy input. Fed-NDIF can also recover false negative lesion regions that are missing or underestimated when using other approaches. For instance, only Fed-NDIF recovered two lesions near the aorta on the shown subject from the Ruijin dataset with a 1\% count level in Fig.\ \ref{comparison}, preserving the shape and quantification of the lesion regions. Similarly, for the Yale subject with a 5\% count level, only DDPET and Fed-NDIF correctly recover the lesion pointed by the arrows. Simultaneously, compared with DDPET, Fed-NDIF removed a false positive region in the lower spine.

Since our goal is to investigate the effectiveness of federated learning on diffusion models for denoising LCPET, Fig.\ \ref{bern}, \ref{ruijin}, and \ref{yale} show the comparison between the local diffusion model DDPET and our federated diffusion model Fed-NDIF for Bern, Ruijin, and Yale datasets with three count levels, respectively. For all datasets and count levels, the error maps indicate substantial reductions in overestimation and underestimation of the predictions, especially in regions with higher contrast, such as the liver, kidney, and brain.

\begin{figure*}[htb!]
    \centering
    \includegraphics[width=\linewidth]{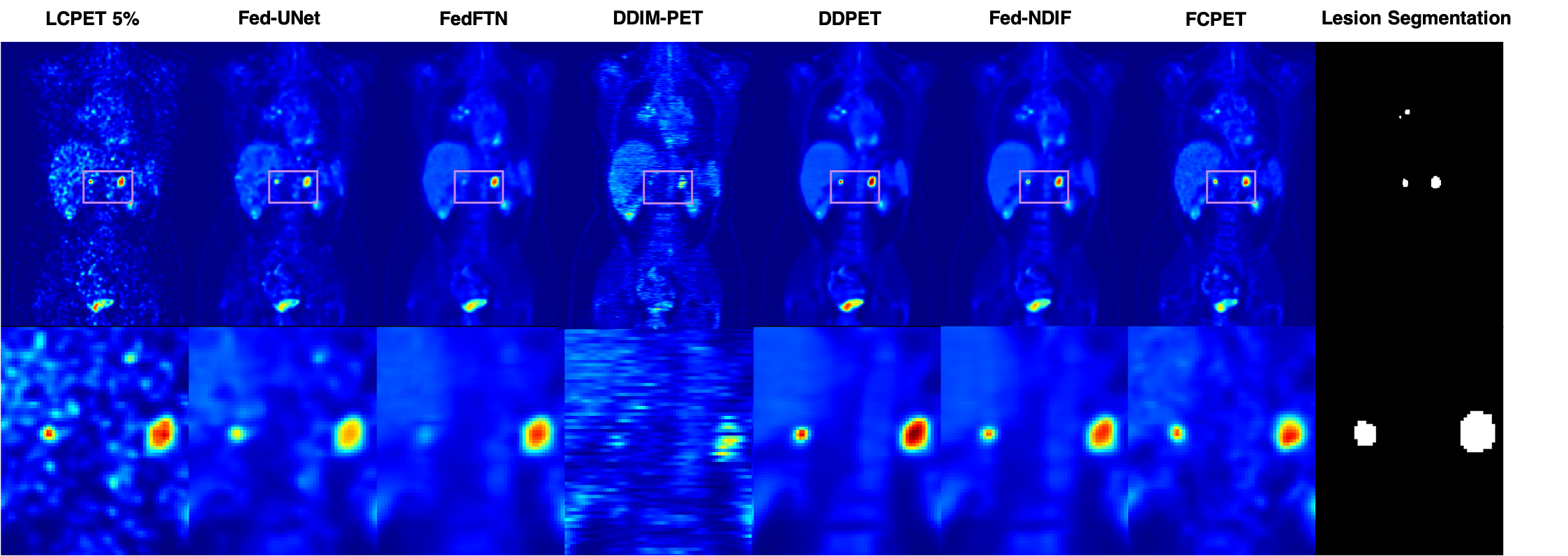}
    \caption{Visualization of tumor quantification at the 5\% count level for Yale data using different methods. The top row shows the whole-body view, and the bottom row displays the magnified lesion region bounded by the purple boxes. The federated diffusion models recover two lesions in the bounding box and reduce the overestimation of the right lesion.}
    \label{tumor}
\end{figure*}

In addition, we also evaluated the accuracy of lesion quantification in the denoised outputs. For the Yale dataset, 15 subjects with 5\%, 10\%, and 20\% count levels with ground truth lesion segmentations were used for evaluation. Table \ref{table2} summarizes the lesion quantification results for federated UNet-based models, local diffusion models, and the proposed Fed-NDIF. Fed-NDIF outperformed all other methods in terms of the percentage error of SUV\textsubscript{mean}, SUV\textsubscript{max}, and NMSE in lesion regions. The diffusion models DDPET and Fed-NDIF achieved better quantification than all federated UNet-based models, while the federated diffusion model further improved the results of DDPET: SUV\textsubscript{mean} error decreased by 2.0\%, SUV\textsubscript{max} error decreased by 3.5\%, and NMSE was reduced by 14.4\%, averaged across all count levels. Fig.\ \ref{tumor} shows the qualitative comparison of lesion quantification using different approaches. For Fed-NDIF, all lesions are identifiable with accurate quantification, while federated UNet-based models and DDIM-PET significantly underestimated the lesions and DDPET overestimated lesion regions.

\subsection{Ablation Studies}

We conducted ablation studies on different components of the proposed framework, including the noise embedding module, the federated global training module, and the fine-tuning module. Table \ref{table3} summarizes the image quality metrics, and Table \ref{table4} summarizes the lesion quantification metrics for the different modules. Firstly, training the local diffusion model with noise embedding improved the baseline model DDPET across nearly all metrics. The addition of federated learning further enhanced image quality across all datasets and count levels. For the image quality of the entire volume, the highest improvement was observed in the Yale dataset, where Fed-NDIF improved NMSE from 0.222 to 0.190 at the 5\% count level, from 0.188 to 0.150 at the 10\% count level, and from 0.159 to 0.117 at the 20\% count level, compared to the local diffusion model with noise embedding (DDPET w/noise). Increases in PSNR and SSIM were also observed in the Yale dataset. Similar improvements in all evaluation metrics were also observed in the Bern and Ruijin datasets. It is important to note that, even without noise embedding, federated learning alone (FED-DDPET) can improve denoising performance compared to local diffusion models for most count levels.
Secondly, for the evaluation of lesion regions shown in table \ref{table4}, Fed-NDIF achieved the lowest SUV\textsubscript{mean} error, SUV\textsubscript{max} error, and NMSE, compared to the local diffusion model (DDPET), the local diffusion model with noise embedding (DDPET w/noise), and the federated diffusion model without noise embedding (Fed-DDPET), underscoring the necessity of the noise embedding and the federated learning modules. Notably, although Fed-DDPET exhibited similar or slightly higher PSNR at a few count levels, Fed-NDIF achieved better tumor quantification than Fed-DDPET across all count levels and metrics. This is crucial as lesion conspicuity is the most important characteristic for radiologists to detect lesions from PET scans and make diagnoses.
Thirdly, we performed ablation studies on Fed-NDIF without the local fine-tuning stage from the federated global model. As shown in Table \ref{table3} and Table \ref{table4}, the absence of fine-tuning (Fed-NDIF w/o ft) led to a significant drop in performance, resulting in even lower image quality and decreased lesion quantification results compared to the local diffusion models. 
Lastly, we tested the number of aggregation steps in the federated global training stage, where local models were trained for a certain number of steps and the model parameters were averaged. We calculated NMSE, SSIM, and PSNR of the denoised outputs with the lowest count level for each dataset using a different number of local training steps: 5000 (5k), 10000 (10k), 20000 (20k), 30000 (30k), and 80000 (80k). Fed-NDIF produced images with the highest quality and accuracy when aggregating local model parameters every 10k steps. As the number of aggregation steps decreased, the performance worsened, with NMSE gradually increasing and PSNR gradually decreasing.

\begin{table*}[htb!]
    \centering
    \scriptsize
    \caption{Local lesion evaluation metrics for different denoising methods at 5\%, 10\%, and 20\% count levels for Yale data (mean ± std), including 15 subjects with lesion segmentations from clinicians. The best results are bolded.}
    \label{table2}
    \begin{adjustbox}{width=\textwidth}
        \begin{tabularx}{\textwidth}{cc|YYY|YYY|YYY}
            \hline
            \multirow{2}{*}{Data} & \multirow{2}{*}{Method} & \multicolumn{3}{c|}{5\%} & \multicolumn{3}{c|}{10\%} & \multicolumn{3}{c}{20\%} \\ \cline{3-11}
            & & SUV\textsubscript{mean} error(\%)$\downarrow$ & SUV\textsubscript{max} error(\%)$\downarrow$ & NMSE $\downarrow$ &
            SUV\textsubscript{mean} error(\%)$\downarrow$ & SUV\textsubscript{max} error(\%)$\downarrow$ & NMSE $\downarrow$& SUV\textsubscript{mean} error(\%)$\downarrow$ & SUV\textsubscript{max} error(\%)$\downarrow$ & NMSE $\downarrow$\\ \hline
            \multirow{6}{*}{Yale}
            & Fed-UNet & 41.2 ± 6.46 & 39.0 ± 9.31 & .388 ± .058 & 26.8 ± 6.51 & 26.3 ± 7.01 & .271 ± .041 & 25.4 ± 4.42 & 27.5 ± 4.83 & .254 ± .041 \\
            & FedFTN & 35.2 ± 12.8 & 41.0 ± 14.8 & .402 ± .047 & 28.5 ± 10.6 & 32.9 ± 12.3 & .243 ± .095 & 20.6 ± 8.35 & 23.3 ± 10.4 & .183 ± .066 \\ 
            & DDIM-PET & 33.5 ± 13.7 & 31.5 ± 13.6 & .460 ± .117 & 26.0 ± 14.6 & 30.1 ± 14.4 & .405 ± .133 & 23.3 ± 14.1 & 27.9 ± 15.6 & .373 ± .140 \\
            & DDPET & 25.1 ± 11.7 & 29.8 ± 14.0 & .276 ± .075 & 19.8 ± 7.78 & 24.8 ± 8.33 & .206 ± .054 & 15.1 ± 7.81 & 19.5 ± 6.89 & .173 ± .043 \\ 
            & \textbf{Fed-NDIF} & \textbf{23.3 ± 13.5}\textsuperscript{\textdagger} & \textbf{27.5 ± 14.1} & \textbf{.244 ± .086}\textsuperscript{\textdagger} & \textbf{18.1 ± 9.85}\textsuperscript{\textdagger} & \textbf{21.0 ± 9.81} & \textbf{.178 ± .066}\textsuperscript{\textdagger} & \textbf{12.6 ± 8.00}\textsuperscript{\textdagger} & \textbf{15.2 ± 6.59}\textsuperscript{\textdagger} & \textbf{.142 ± .056}\textsuperscript{\textdagger} \\ \hline
        \end{tabularx}
    \end{adjustbox}
    \begin{minipage}{\textwidth}
    \vspace{0.1cm}
    \scriptsize \textdagger \ $p < 0.005$ between DDPET and Fed-NDIF based on Wilcoxon rank test. 
    \end{minipage}
\end{table*}

\begin{table*}[htb!]
    \centering
    \scriptsize
    \caption{Ablation studies on noise embedding and federated learning at three count levels for three datasets (30 subjects, mean)}
    \label{table3}
    \begin{adjustbox}{width=\textwidth}
        \begin{tabularx}{\textwidth}{cc|YYY|YYY|YYY}
            \hline
            \multirow{2}{*}{Data}& \multirow{2}{*}{Method} & \multicolumn{3}{c|}{1\%} & \multicolumn{3}{c|}{5\%} & \multicolumn{3}{c}{10\%} \\ \cline{3-11}
             & & PSNR $\uparrow$ & SSIM $\uparrow$ & NMSE $\downarrow$ & PSNR $\uparrow$ & SSIM $\uparrow$ & NMSE $\downarrow$ & PSNR $\uparrow$ & SSIM $\uparrow$ & NMSE $\downarrow$ \\ \hline
            \multirow{5}{*}{Bern} & DDPET & 55.242  & .979  & .190 & 57.162 & .985 & .152 & 59.305 & .988  & .119  \\ 
             & DDPET w/noise & 55.784 &  .983  &  .182  &  58.958  & .989  & .124  &  60.233 & .989  & .107 \\ 
             & Fed-DDPET & 55.782 & .982 & .180 & 59.265 & .989 & \textbf{.120} & \textbf{60.628} & .989 & \textbf{.103} \\ 
             & Fed-NDIF w/o ft & 54.064 & .969 & .218 & 56.705 & .982 & .159 & 56.765 & .981 & .158 \\
             & \textbf{Fed-NDIF} & \textbf{55.799} & \textbf{.983} & \textbf{.179 } & \textbf{59.266} & .989 & \textbf{.120} & 60.615 & .989 & \textbf{.103} \\ \hline \hline
             \multirow{2}{*}{Data}& \multirow{2}{*}{Method} & \multicolumn{3}{c|}{1\%} & \multicolumn{3}{c|}{5\%} & \multicolumn{3}{c}{10\%} \\ \cline{3-11}
             & & PSNR $\uparrow$ & SSIM $\uparrow$ & NMSE $\downarrow$ & PSNR $\uparrow$ & SSIM $\uparrow$& NMSE $\downarrow$ & PSNR $\uparrow$ & SSIM $\uparrow$ & NMSE $\downarrow$ \\ \hline
            \multirow{5}{*}{Ruijin} & DDPET & 50.796  & .968 & .226  & 55.306& .983 & .136  & 56.896& .984 & .114 \\
             & DDPET w/noise & 51.207 &  .967  &  .216  &  55.534  & .983  & .134  &  57.112 & .985  & .112 \\ 
             & Fed-DDPET & 51.146 & .968 & .217 & 55.723 & .983 & .131 & \textbf{57.496} & .985 & \textbf{.109} \\
             & Fed-NDIF w/o ft & 49.283 & .943 & .270 & 53.610 & .971 & .164 & 54.561 & .976 & .147 \\
             & \textbf{Fed-NDIF} & \textbf{51.316} & \textbf{.970} & \textbf{.214} & \textbf{55.786} & \textbf{.984} & \textbf{.130} & 57.264 & .985 & \textbf{.109} \\ \hline \hline
             \multirow{2}{*}{Data}& \multirow{2}{*}{Method}& \multicolumn{3}{c|}{5\%} & \multicolumn{3}{c|}{10\%} & \multicolumn{3}{c}{20\%} \\ \cline{3-11}
             & & PSNR $\uparrow$ & SSIM $\uparrow$ & NMSE $\downarrow$ & PSNR $\uparrow$ & SSIM $\uparrow$& NMSE $\downarrow$ & PSNR $\uparrow$ & SSIM $\uparrow$ & NMSE $\downarrow$ \\ \hline
            \multirow{5}{*}{Yale} & DDPET & 52.600 & .957 & .231 & 54.085 & .968& .194 &  55.418& .975  & .167  \\ 
             & DDPET w/noise &  52.920  &  .957  &  .222  & 54.345  & .967  & .188 &  55.804  & .975 & .159 \\
             & Fed-DDPET & \textbf{54.341} & .965 & \textbf{.190} & 56.327 & .974 & .151 & 58.506 & .980 & .118 \\
             & Fed-NDIF w/o ft & 52.330 & .948 & .238 & 54.400 & .966 & .186 & 55.616 & .973 & .161 \\
             & \textbf{Fed-NDIF} & 54.337 & \textbf{.968} & \textbf{.190} & \textbf{56.364} & \textbf{.977} & \textbf{.150} & \textbf{58.547} & \textbf{.984} & \textbf{.117} \\ \hline
        \end{tabularx}
    \end{adjustbox}
\end{table*}

\begin{table*}[htb!]
    \centering
    \scriptsize
    \caption{Ablation studies on noise embedding and federated learning for local lesion evaluation at three count levels on the Yale mCT dataset (15 subjects with lesion segmentation)}
    \label{table4}
    \begin{adjustbox}{width=\textwidth}
        \begin{tabularx}{\textwidth}{cc|YYY|YYY|YYY}
            \hline
            \multirow{2}{*}{Data} & \multirow{2}{*}{Method} & \multicolumn{3}{c|}{5\%} & \multicolumn{3}{c|}{10\%} & \multicolumn{3}{c}{20\%} \\ \cline{3-11}
            & & SUV\textsubscript{mean} error(\%)$\downarrow$ & SUV\textsubscript{max} error(\%)$\downarrow$ & NMSE $\downarrow$ &
            SUV\textsubscript{mean} error(\%)$\downarrow$ & SUV\textsubscript{max} error(\%)$\downarrow$ & NMSE $\downarrow$& SUV\textsubscript{mean} error(\%)$\downarrow$ & SUV\textsubscript{max} error(\%)$\downarrow$ & NMSE $\downarrow$\\ \hline
            \multirow{5}{*}{Yale}
            & DDPET & 25.1 & 29.8 & .276 & 19.8 & 24.8 & .206 & 15.1 & 19.5 & .173 \\ 
            & DDPET w/noise & 24.0 & 30.8 & .269 & 19.6 & 24.6 & .207 & 14.9 & 19.5 & .174 \\ 
            & Fed-DDPET & 24.9 & 28.5 & .250 & 19.1 & 21.4 & .181 & 13.3 & 16.1 & \textbf{.142} \\ 
            & Fed-NDIF w/o ft & 24.4 & 29.0 & .252 & 18.5 & 22.1 & .200 & 16.0 & 21.7 & .188 \\ 
            & \textbf{Fed-NDIF} & \textbf{23.3} & \textbf{27.5} & \textbf{.244} & \textbf{18.1} & \textbf{21.0} & \textbf{.178} & \textbf{12.6} & \textbf{15.2} & \textbf{.142} \\ \hline
        \end{tabularx}
    \end{adjustbox}
\end{table*}

\begin{figure*}[htb!]
    \centering
    \includegraphics[width=\linewidth]{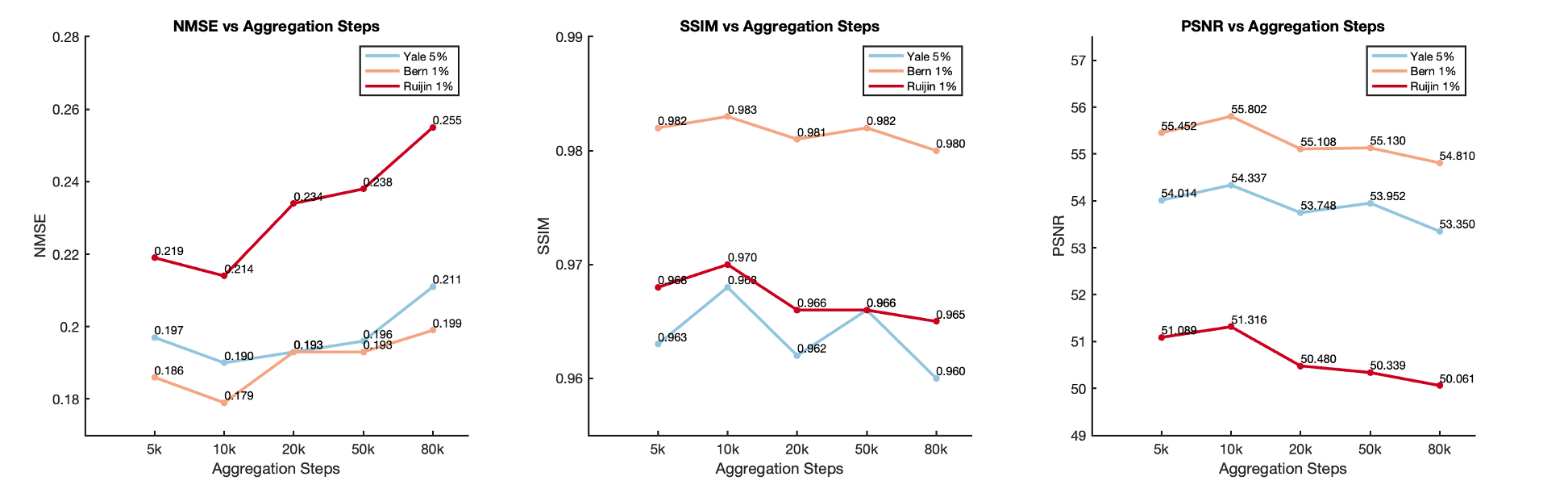}
    \caption{Ablation studies on the number of aggregation steps for subjects with the lowest count level from the Bern, Ruijin, and Yale datasets. The local training steps of 10k between each aggregation achieved the best results for all datasets.}
    \label{ablation}
\end{figure*}


\section{Discussion}
Our work proposes a federated learning framework, Fed-NDIF, for LCPET denoising with diffusion models. Given that diffusion models have achieved excellent denoising performance at extremely low count levels, federated learning could further improve model performance by taking advantage of model parameters trained using data from additional institutions. This is crucial when an institution only has limited data, as diffusion models might memorize the training data and exhibit low generalizability to new testing data, especially when the training data are not diverse enough \cite{somepalli2023diffusion, gu2023memorization}. The largest improvements in NMSE and PSNR were observed on the Yale and Bern datasets, which had fewer training data than the Ruijin dataset. Additionally, as indicated in Table \ref{table3}, embedding image noise into the diffusion model can enhance model performance in both scenarios: when training the model locally and when training the model in a federated learning approach. For local training, noise embedding makes the diffusion model noise-aware, allowing it to identify inputs from different count levels and generalize to all count levels and injected doses. Many studies have shown the impact of noise levels in training and testing images, and denoised outputs are most desirable when training and testing images have similar noise levels \cite{liu2021impact, li2022noise}. However, training images with multiple count levels can increase the size of the training data. Additionally, one model that generalizes to different count levels can avoid the need to train multiple models, which is more time-consuming. In federated learning, image noise variation across datasets can pose additional challenges for model denoising performance if the data are non-IID (independent and identically distributed) or heterogeneous \cite{zhu2021federated, zhao2018federated}. In a federated diffusion model, image noise can be used as a standardized metric to denote data variation across institutions and map data distribution. As shown in Table \ref{table4}, adding noise embedding to the federated diffusion model can improve lesion quantification in terms of SUV\textsubscript{mean}, SUV\textsubscript{max}, and NMSE at all count levels for the Yale dataset, compared to the federated diffusion model without noise embedding. Combining federated learning and noise embedding with the diffusion model for PET denoising, our proposed Fed-NDIF can produce images with higher quality and more accurate lesion quantification, even reducing false positive and false negative predictions, which are important for clinicians to precisely localize pathology regions and make diagnoses. 

We acknowledge that limitations still exist for our proposed method. Training time and the cost of federated diffusion models, especially due to the heavy communication overhead in the federated learning stage, remain concerns. As shown in the ablation studies of the aggregation steps (Fig.\ \ref{ablation}), model performance increases with fewer local training steps and more communication rounds; however, training time and the number of parameters needed to be averaged also increase linearly \cite{de2024training, jothiraj2023phoenix}, especially as the number of institutions increases. Strategies to boost communication efficiency, such as averaging model parameters partially or distilling knowledge from local models, could be implemented to reduce the communication burden \cite{de2024training, wu2022communication}. Additionally, our diffusion model is implemented in 2.5D due to limitations in computational resources. However, the proposed framework is flexible and can be extended to a 3D diffusion model, which might produce even better denoising results \cite{yu2024pet, xia2024anatomically}.

\section{Conclusion}
In our study, we introduce a noise-embedded federated learning diffusion model (Fed-NDIF) for the denoising of low-count PET images. Our approach effectively integrates liver normalized standard deviation as an image noise embedding within a 2.5D conditional diffusion model and employs the Federated Averaging algorithm to aggregate model parameters across multiple institutions without compromising data privacy. Extensive validation on datasets from the University of Bern, Ruijin Hospital in Shanghai, and Yale-New Haven Hospital, encompassing multiple count levels, demonstrates the superiority of our method in enhancing image quality and improving lesion quantification compared to federated UNet-based and local diffusion models. Through comprehensive ablation studies, we emphasize the importance of noise embedding and federated learning modules, confirming the robustness of Fed-NDIF in producing high-quality denoised outputs and achieving accurate tumor quantification. By preserving data privacy and addressing the heterogeneity in noise levels, our Fed-NDIF method demonstrates the potential of federated learning to advance diffusion models for LCPET image denoising.
\section*{Acknowledgments}
This work was supported by the National Institutes of Health (NIH) grants R01EB025468 and R01CA275188.

\section*{Declaration of Competing Interest}
The authors declare that they have no known competing financial interests or personal relationships that could have appeared to influence the work reported in this paper.


\bibliography{reference}
\bibliographystyle{IEEEtran}

\end{document}